\documentclass[twocolumn,amsmath,amssymb,aps,prl,floatfix,showpacs,citeautoscript]{revtex4}

\usepackage{epsfig}
\usepackage{colordvi}
\usepackage{graphicx}

\begin{document}

\def\bea{\begin{eqnarray}}
\def\nn{\nonumber\\}
\def\eea{\end{eqnarray}}
\def\beq{\begin{equation}}
\def\eeq{\end{equation}}

\def\kl{\sigma^a}
\def\klb{{\boldsymbol\sigma}^a}
\def\M{{\bf M}}
\def\J{{\bf J}}
\def\rhoah{\rho^a}
\def\bsig{{\boldsymbol\sigma}}
\def\balpha{\hat{\boldsymbol\alpha}}
\def\bbeta{\hat{\boldsymbol\beta}}
\def\bgamma{\hat{\boldsymbol\gamma}}
\def\poly{\kl_{\rm poly}}
\def\mdir{{\bf\hat m}}
\def\th{\hat{\boldsymbol\theta}}
\def\ph{\hat{\boldsymbol\varphi}}
\def\x{\hat{\bf x}}
\def\y{\hat{\bf y}}
\def\z{\hat{\bf z}}
\def\sigpar{\sigma_m}
\def\sigperp{\sigma_\theta}
\def\sigperpperp{\sigma_\varphi}

\title{Orientation Dependence of the Intrinsic Anomalous Hall Effect in hcp Cobalt}

\author{Eric Roman}
\affiliation{Department of Physics, University of California,
Berkeley, CA 94720, USA}
\author{Yuriy Mokrousov}
\affiliation{Department of Physics, University of California,
Berkeley, CA 94720, USA}
\author{Ivo Souza}
\affiliation{Department of Physics, University of California,
Berkeley, CA 94720, USA}

\date{\today}

\begin{abstract}
  We carry out first-principles calculations of the dependence of the
  intrinsic anomalous Hall conductivity vector $\klb$ of hcp Co on the
  magnetization direction $\mdir$.  The magnitude of $\klb$ decreases
  smoothly from $481$~S/cm to $116$~S/cm as the magnetization is
  tilted from the $c$-axis to the $ab$-plane.  This factor-of-four
  reduction compares well with measurements on single crystals, while
  the angular average $\langle\klb\cdot\mdir\rangle=226$~S/cm is in
  excellent agreement with the value of 205~S/cm measured in
  polycrystalline films.  The strong anisotropy of $\klb$ is a
  consequence of spin-orbit induced changes in the electron states
  near the Fermi level as the magnetization is rotated.
\end{abstract}

\maketitle


The Hall effect in ferromagnets has two contributions. The ordinary
Lorentz-force part, which is proportional to the magnetic field for
weak fields, and an ``anomalous'' part, which depends on the
magnetization.  A theory of the anomalous Hall effect (AHE) was put
forth by Karplus and Luttinger \cite{karplus54}, who showed that a
Hall current perpendicular to the electric field and odd under
magnetization reversal is established in a ferromagnetic crystal as a
result of the spin-orbit interaction (SOI).

The link between the AHE and the SOI became consensual, but a long
debate ensued on whether the relevant SOI is associated with the
crystalline potential (intrinsic) or with impurity atoms (extrinsic).
The asymmetric impurity scattering of the spin-polarized charge
carriers (skew-scattering) leads to a linear dependence of the
anomalous Hall resistivity $\rhoah_{yx}=E_y/j_x$ on the longitudinal
resistivity $\rho_{xx}$; an additional scattering process, side-jump,
yields the same scaling $\rhoah_{yx}\propto\rho_{xx}^2$ as the
intrinsic Karplus-Luttinger mechanism.  For a review see
Ref.~\onlinecite{nagaosa09}.

Unlike the extrinsic contributions, which depend on the details of the
impurity potential, the Karplus-Luttinger anomalous Hall conductivity
(AHC) can be evaluated from the band structure of the crystal
\cite{nagaosa09,yao04}:
\beq
\label{eq:ahc}
\kl_{ij}=-\frac{e^2}{\hbar}\int_{\rm BZ} \frac{d^3k}{(2\pi)^3}\sum_n
f_{n\bf k}\Omega_{n{\bf k},ij}, \eeq
where $f_{n\bf k}$ is the Fermi-Dirac distribution function and
$\Omega_{n,ij}({\bf k})$ is the Berry curvature tensor of each
cell-periodic spinor Bloch state $|u_{n\bf k}\rangle$.
First-principles calculations for Fe, Co, Ni
\cite{yao04,wang06,wang07}, SrRuO$_3$ \cite{fang03} and
Mn$_5$Ge$_3$\cite{zeng06} have consistently found good agreement with
room-temperature experiments, establishing the importance of the
intrinsic mechanism.

Recent experiments have focused on isolating the different
contributions to the AHE \cite{kotzler05,zeng06,tian09}.
Skew-scattering can be separated from the other two terms by fitting
the measured anomalous Hall resistivity to the form
\beq
\label{eq:contribs}
\rhoah_{yx}=a\rho_{xx}+b\rho_{xx}^2,
\eeq
where $b=\kl_{xy}+b^{\rm SJ}$.  The coefficients $a$ (skew-scattering)
and $b$ (intrinsic plus side-jump) can be read off a plot of
$\rhoah_{yx}/\rho_{xx}$ versus $\rho_{xx}$, where $\rho_{xx}$ is
varied through doping or temperature changes.

In this Letter, we present a detailed first-principles study of the
orientation dependence of the intrinsic AHE.  Anisotropy in the AHE
has been measured in single crystals of the ferromagnetic elements
(hcp Co \cite{volkenshtein61}, hcp Gd \cite{lee67}, fcc Ni
\cite{hiraoka68}, and bcc Fe \cite{hirsch73}), and more recently in
ferromagnetic compounds \cite{ohgushi06,sales08,stankiewicz08}.  On
the theoretical side there has been little progress beyond the basic
phenomenological description. To our knowledge, the only attempt at a
microscopic model has been the tight-binding study of
Ref.~\onlinecite{ohgushi06}. Because the AHC is very sensitive to fine
details in the band structure \cite{yao04,fang03,wang06}, a
quantitative {\it ab initio} theory of anisotropy is highly desirable.
It is also not obvious that the phenomenological description of
magnetocrystalline anisotropy \cite{birss64,hurd74} applies to the AHC
given by Eq.~(\ref{eq:ahc}).  The Berry curvature undergoes strong and
rapid variations in $k$-space, with sharp peaks and valleys from
avoided crossings near the Fermi level \cite{yao04,fang03,wang06}.  It
has been argued that such behavior cannot be described perturbatively
\cite{yao04}, and that it often gives rise to a complex or even
irregular behavior of the AHC as a function of exchange splitting and
Fermi level position \cite{fang03}.  This raises the possibility that
the orientation dependence of $\kl_{ij}$ may also not be smooth. We
find instead that in hcp Co it is strong but remarkably smooth, and
can be described by a phenomenological power-series expansion.  The
calculated anisotropy accounts for the experimental observations in
both single crystals (angular dependence) and polycrystalline films
(angular average).

We begin by reviewing the phenomenology
\cite{birss64,hurd74}. Electrical conduction in ferromagnets is
described by a magnetization-dependent conductivity tensor:
$J_i=\sigma_{ij}(\M)E_j$, where $i,j$ are Cartesian indices.  The
Onsager relation $\sigma_{ij}(\M)=\sigma_{ji}(-\M)$ implies that the
symmetric ($\sigma_{ij}^s$) and antisymmtric ($\sigma_{ij}^a$) parts
of $\sigma_{ij}$ are respectively even and odd functions of the
magnetization $\M$.  The current density $\J$ is therefore comprised
of an even (Ohmic) part $\J^s$ and an odd (Hall) part $\J^a$.  The
Hall current reads $\J^a={\bf E}\times\bsig^a$, where
$\sigma^a_k=(1/2)\epsilon_{ijk}\sigma_{ij}^a$.  $\J^a$ is
perpendicular to ${\bf E}$ but not necessarily to $\M$, since
$\bsig^a$ and $\M$ may not be collinear.  This is the situation in
single crystals, where $\bsig^a\parallel\M$ only when $\M$ points
along certain high-symmetry directions.

To characterize the anisotropy we write $\klb$ as
\beq
\label{eq:sigma-decomp}
\klb=\sigpar\mdir+\sigperp\th+\sigperpperp\ph,
\eeq
where $\theta$ and $\varphi$ are respectively the polar and azimuthal
angles of $\M=M\mdir$. In macroscopically isotropic systems such as
polycrystals $\klb\parallel\M$, i.e., $\sigperp=\sigperpperp=0$.  The
orientation dependence in hcp single crystals is given to third order
in a spherical-harmonic expansion by \cite{suppl}
\beq
\label{eq:expansion-ylm}
\begin{cases}
\sigma_1^a=c_{11}\overline{Y}_{11}+c_{31}\overline{Y}_{31}\\
\sigma_2^a=c_{11}\overline{Y}_{1,-1}+c_{31}\overline{Y}_{3,-1}\\
\sigma_3^a=c_{10}\overline{Y}_{10}+c_{30}\overline{Y}_{30}
\end{cases},
\eeq
where $\overline{Y}_{lm}(\theta,\varphi)$ are real spherical
harmonics.  Because $\sigma_3^a$ is independent of $\varphi$, while
$\sigma_1^a$ and $\sigma_2^a$ have respectively cosine and sine
dependences, $\klb$ and $\M$ share the same azimuthal angle, and their
polar-angle mismatch is independent of $\varphi$. Thus
$\sigma_{m,\theta}=\sigma_{m,\theta}(\theta)$ and $\sigperpperp=0$.

\begin{figure}
\begin{center}
\epsfig{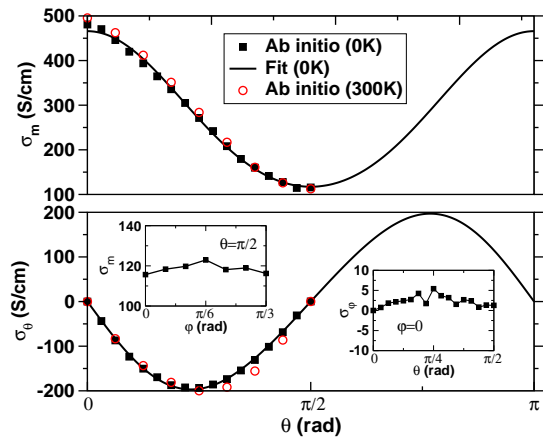}
\end{center}
\caption{(Color online.) Evolution of the components of the anomalous
  Hall conductivity parallel ($\sigpar$) and perpendicular
  ($\sigperp$) to the magnetization [Eq.~(\ref{eq:sigma-decomp})] as
  $\M$ is tilted by $\theta$ from the $c$-axis towards the $a$-axis.
  The solid lines are fits to the first-principles data, as described
  in the text. The left and right insets shows respectively
  $\sigpar(\pi/2,\varphi)$, and $\sigperpperp(\theta,0)$.  }
\label{fig:parperp}
\end{figure}

We have carried out fully-relativistic band-strucutre calculations for
hcp and fcc Co at the experimental lattice constants of $4.74$ and
$6.68$ bohr respectively, using the {\tt PWSCF} code \cite{pwscf}.
The pseudopotential was generated using similar parameters as in
Ref.~\onlinecite{wang06}. The plane-wave basis cutoff was set at
140~Ry, and the exchange-correlation functional, $k$-point sampling,
and Fermi smearing used in the self-consistent calculation are the
same as in Ref.~\onlinecite{wang06}. The integral in
Eq.~(\ref{eq:ahc}) was evaluated using a Wannier-interpolation scheme
\cite{wang06} to sample efficiently the Brillouin zone over a
$125\times 125\times 125$ uniform $k$-point mesh ($200\times 200\times
200$ for fcc Co), with a $5\times 5\times 5$ adaptively refined mesh
around the points where the magnitude of the Berry curvature exceeded
10~\AA$^2$.

The AHC of hcp Co was calculated for several orientations of the
cell-averaged magnetization \cite{explan-cellavg} in the $ac$-plane
($\varphi=0$).  The tilting angle $\theta$ was increased from $0$
($\M\parallel c$-axis) to $\pi/2$ ($\M\parallel a$-axis) in steps of
$\pi/32$, and for each step the vector $\klb(\theta,\varphi)$ was
calculated.  Fig.~\ref{fig:parperp} contains the numerical results:
$\sigma_m(\theta,0)$ and $\sigma_\theta(\theta,0)$ are shown in the
main panels, while the insets contain additional data which confirms
the absence of (or very weak) basal-plane anisotropy.  The vectors
$\klb$ and $\M$ start out parallel, but as $\M$ begins to tilt away
from the $c$-axis $\klb$ lags behind ($\sigma_\theta<0$), and they
become parallel again upon reaching the basal plane.  The AHC is
strongly anisotropic, decreasing in magnitude by a factor of
$481/116\simeq 4.1$ between $\theta=0$ and $\theta=\pi/2$. This is in
reasonable agreement with the ratio of 2.93 measured in single
crystals at 290~K \cite{volkenshtein61}.  While strong, the angular
dependence of $\klb$ is smooth, and can be described by
Eq.~(\ref{eq:expansion-ylm}). A least-squares fitting to the data
yields, in S/cm, $c_{10}=951.5$, $c_{11}=-204.1$, $c_{30}=1.2$, and
$c_{31}=38.4$, producing the solid-line curves in
Fig.~\ref{fig:parperp}.

\begin{table}
  \caption{
    Anomalous Hall conductivity $|\klb|=\sigma_m$ in S/cm  for selected 
    high-symmetry orientations of the magnetization in hcp and fcc Co.
    The AHC of polycrystalline samples is calculated as an orientational
    average (see text).}
\begin{ruledtabular}
\begin{tabular}{cccc}
Co & Orientation & Calc. &Expt.\cr
\hline
hcp & $c$-axis   & 481                    & 683\footnotemark[1]\cr 
    & $ab$-plane & 116 & 232\footnotemark[1]\cr
    & Polycrystal & 226 & 205\footnotemark[2],\,275\footnotemark[3]\cr 
\hline
fcc & [001] & 249& \cr
    & [110] & 218& \cr
    & [111] & 234& \cr
\end{tabular}
\end{ruledtabular}
\footnotetext[1]{Reference~\onlinecite{volkenshtein61}} 
\footnotetext[2]{Reference~\onlinecite{kotzler05}} 
\footnotetext[3]{Reference~\onlinecite{volkenshtein60}}
\label{table:results}
\end{table}

We now turn to the comparison with the measurements on polycrystalline
films \cite{kotzler05}.  The films were magnetized along the growth
direction by an applied field; assuming randomly oriented
crystallites, each with a bulk-like Hall current density
$\J^a(\theta,\varphi)$, the net Hall current density in the films can
be estimated by performing an orientational average:
$\langle\J^a\rangle={\bf E}\times\langle\bsig^a\rangle$.  Because
$\klb$ displays azimuthal isotropy, it suffices to average
Eq.~(\ref{eq:sigma-decomp}) over $\theta$ for fixed $\varphi$. The
average of $\sigperp$ vanishes (see Fig.~\ref{fig:parperp}), resulting
in an isotropic AHC $\langle\klb\rangle=\langle\sigpar\rangle\mdir$ of
magnitude $\poly=
\int_0^{\pi/2}\,\sigpar(\theta)\sin\theta\,d\theta=226$~S/cm.  This
should be compared with the value $b=205$~S/cm obtained in
Ref.~\onlinecite{kotzler05} by fitting the experimental data to
Eq.~(\ref{eq:contribs}).  The experiment does not discriminate between
the intrinsic and side-jump components of $b$, but the close agreement
with the calculated $\poly$ reinforces the conclusion \cite{kotzler05}
that the former dominates.  Table~\ref{table:results} summarizes the
comparison between our calculations and experiments.

\begin{figure}
\begin{center}
\epsfig{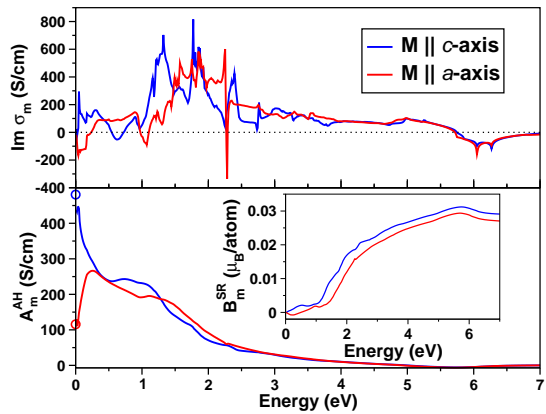}
\end{center}
\caption{(Color online.) Upper panel: MCD spectrum for two
  magnetization directions. Lower panel: Cumulative contribution to the
  AHC from the spectrum above energy $\hbar\omega$. The circles
  denote the AHC calculated directly from Eq.~(\ref{eq:ahc}). Inset:
  Cumulative contribution to the self-rotation part of the orbital
  magnetization from the spectrum below $\hbar\omega$.}
\label{fig:mcd}
\end{figure}

Next we discuss the origin of the strong anisotropy.  The AHE of
uniaxial crystals is anisotropic to first order in an expansion in
powers of the magnetization \cite{suppl}, while in cubic crystals
anisotropy appears only in third-order \cite{hiraoka68}, and is
expected to be much weaker.  For example, the AHC of fcc Co changes by
less than 10\% as a function of the magnetization direction
(Table~\ref{table:results}).  Perhaps more surprising is the fact that
the AHE in hcp Co appears to be considerably more anisotropic than
both the magneto-optical spectrum \cite{weller94} and the orbital
magnetization \cite{trygg95,ceresoli09}. This is intriguing because
the three phenomena are related by linear sum rules \cite{souza08},
and hence anisotropy appears at the same order.

The sum rules read $\langle\omega^{-1}{\rm
  Im}\,\klb\rangle_\omega=(\pi/2)\klb(\omega=0)$ and $\langle {\rm
  Im}\,\klb\rangle_\omega=(\pi ec/\hbar)\M_{\rm SR}^{\rm(I)}$, where
$\langle f\rangle_\omega=\int_0^\infty f(\omega)d\omega$.
$\klb(\omega=0)$ is the dc AHC; at finite frequencies $\klb$ acquires
an imaginary part which describes the differential absorption of right
and left circularly-polarized light, or magnetic circular dichroism
(MCD). The first sum rule expresses the AHC in terms of the first
inverse moment of the MCD spectrum. The second relates $\M_{\rm
  SR}^{\rm(I)}$, the ``gauge-invariant self-rotation'' part of the
orbital magnetization \cite{souza08}, to the zero-th spectral moment.

\begin{figure}
\begin{center}
\epsfig{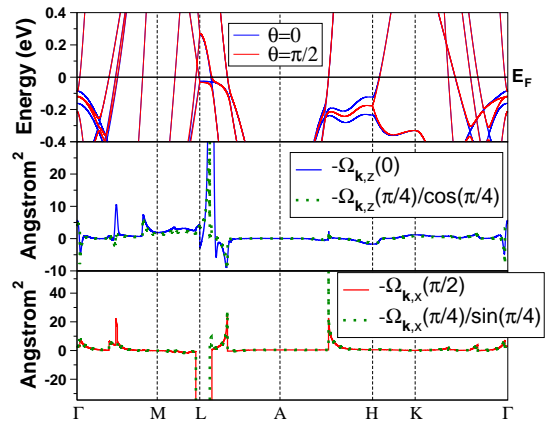}
\end{center}
\caption{(Color online.) Upper panel: Energy bands close to the Fermi
  level for $\M\parallel$ $c$-axis ($\theta=0$) and $\M\parallel$
  $a$-axis ($\theta=\pi/2$). Middle and lower panels: $k$-space Berry
  curvature ${\boldsymbol\Omega}_{\bf k}(\theta)$ for $\theta=0$,
  $\theta=\pi/2$, and $\theta=\pi/4$. The full height of the
resonance peaks near the $L$-point is of the order of $10^3$~\AA$^2$.}
\label{fig:kpath}
\end{figure}

The absorptive part of $\sigma_m(\omega)$ is plotted in the upper
panel of Fig.~\ref{fig:mcd} for $\theta=0,\pi/2$.  The lower panel
shows $A_m^{\rm AH}(\omega)=\frac{2}{\pi} \int_\omega^{\omega_{\rm
    max}}\,\frac{1}{\omega'} {\rm Im}\,\sigpar(\omega')\,d\omega'$.
For $\omega_{\rm max}\rightarrow\infty$ (in practice we 
use $\omega_{\rm max}=7$~eV) $A_m^{\rm AH}(0)=\sigpar(\omega=0)$, so
that $A_m^{\rm AH}(\omega>0)$ is the cumulative contribution to the
AHC from optical transitions above $\omega$.  While for either
orientation there are sizeable contributions to the AHC up to
$\omega\sim 3.5$~eV, its orientation dependence is concentrated below
0.3~eV.  At these low frequencies the MCD spectrum changes sign
between $\theta=0$ and $\theta=\pi/2$. This difference gets magnified
in the AHC via the inverse-frequency weight factor, producing the
bifurcation below 0.3~eV of the two $A_m^{\rm AH}(\omega)$ curves.
All frequencies are equally weighted in the orbital moment sum rule,
which as a result is more isotropic.  This is seen in the inset, where
we plot $B_m^{\rm SR}(\omega)=\frac{V_c\hbar}{2\pi ec}
\int_0^\omega{\rm Im}\,\kl_m(\omega')d\omega'$, the cumulative
contribution below $\omega$ to the gauge-invariant self-rotation per
atom.

The origin of the low-frequency anisotropy can be seen in
Fig.~\ref{fig:kpath}.  The upper panel displays the energy bands for
the two magnetization directions.  Rotating $\M$ from the $c$-axis to
the $a$-axis in the presence of the SOI turns various band crossings
into avoided crossings and vice-versa. When this occurs close to the
Fermi level the Berry curvature along $\M$ can flip sign in the
process, while retaining a large magnitude.  This is what happens near
the $L$-point, as seen in the middle and lower panels, where we plot
$-{\boldsymbol \Omega}_{\bf k}\cdot\mdir$ for the two orientations
($\Omega_{{\bf k},k}=(1/2)\epsilon_{ijk} \sum_n f_{n\bf
  k}\Omega_{n{\bf k},ij}$ is the total Berry curvature at ${\bf k}$).
The sensitivity of the AHC to changes in the electron states near
$E_F$ may also be understood from the fact that $\klb$ can be recast
as a Fermi surface integral \cite{wang07,haldane04}, whereas the
orbital magnetization truly depends on all occupied states.
 
How can the spiky behavior of ${\boldsymbol\Omega}_{\bf k}$ be
reconciled with the smooth angular dependence displayed by
$\klb(\theta)$ in Fig.~\ref{fig:parperp}? According to the
phenomenological expansion (\ref{eq:expansion-ylm}), $\kl_i\propto
M_i$ to leading order ($i=x,y,z$).  This will be the case for the AHC
given by Eq.~(\ref{eq:ahc}) provided that $\Omega_{{\bf k},i}\propto
M_i$ at each ${\bf k}$.  This proportionality holds reasonably well
even around strong resonance peaks, judging from the comparison in
Fig.~\ref{fig:kpath} between ${\boldsymbol\Omega}_{\bf k}$ calculated
at $\theta=\pi/4$ and at $\theta=0,\pi/2$.

\begin{figure}
\begin{center}
\epsfig{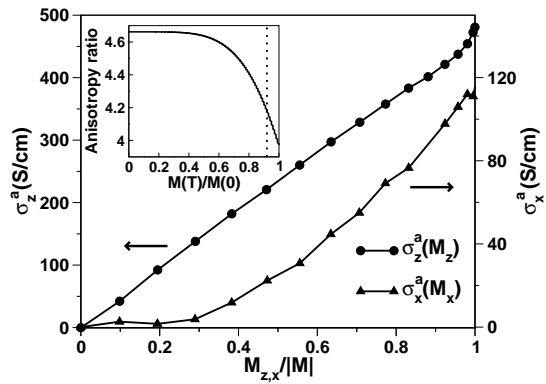}
\end{center}
\caption{Evolution of $\klb$ as the magnetization is rotated in the
  $ac$ plane, plotted as $\kl_z(\cos\theta)$ and
  $\kl_x(\sin\theta)$. Inset: Anisotropy ratio
  $\kl_z(\theta=0)/\kl_x(\theta=\pi/2)$ versus the reduced
  magnetization, according to the spin-fluctuation model.  The dotted
  line denotes the approximate location of the hcp$\rightarrow$fcc
  transition.}
\label{fig:sigma-M}
\end{figure}

We end with a discussion of temperature effects. The AHC hardly
changes as the Fermi-smearing temperature in Eq.~(\ref{eq:ahc}) is
varied from 0~K to 300~K (Fig.~\ref{fig:parperp}).  This agrees with
the constancy of the coefficient $b$ of polycrystalline films in the
range 78-350~K \cite{kotzler05}.  The angular dependence of the AHC
can also give rise to a temperature dependence, via long-wavelength
thermal fluctuations in $\mdir$ \cite{zeng06}. According to this
model, if $\kl_i(T=0)$ changes linearly with $M_i$ upon rotating $\M$
then $\klb(T)=[M(T)/M(0)]\klb(T=0)$, which seems to explain the
temperature dependence in Mn$_5$Ge$_3$ \cite{zeng06}.
Fig.~\ref{fig:sigma-M} shows that in hcp Co $\kl_z$ depends linearly
on $M_z$, while the $\kl_x(M_x)$ curve is significantly nonlinear
(note that $|c_{31}/c_{11}|\gg |c_{30}/c_{10}|$).  As a result, the
$a$-axis AHC should decrease with $T$ faster than $M(T)$, producing an
increase with temperature of the ratio
$\kl_z(\theta=0)/\kl_x(\theta=\pi/2)$.  An estimate of the magnitude
of this effect can be obtained using the ``$l(l+1)/2$ power law''
\cite{zener54,callen66} for the coefficients $c_{lm}(T)$ (see
Ref.~\onlinecite{suppl} for details).  The result, shown in the inset
of Fig.~\ref{fig:sigma-M}, is a 17\% increase between 0~K and
$T_c$. This is a sufficiently large effect that is should be
observable in principle, but in practice it is preempted by the phase
transformation into the fcc structure at 695~K, well below
$T_c=1400$~K \cite{explan-tsoukalas}.

In summary, we have shown by means of first-principles calculations
that the intrinsic mechanism for the AHE describes quantitatively the
observed strong angular dependence in hcp Co single crystals. The key
role of near-degeneracies across the Fermi level was elucidated, and
the AHE of polycrystalline Co films was reproduced by averaging the
single-crystal Hall conductivity over all magnetization directions.
Further experimental and theoretical studies of the orientation
dependence of the AHE are needed. For example, very little is known
about the anisotropy of the skew-scattering contribution
\cite{stankiewicz08}.

The authors wish to thank David Vanderbilt, Jonathan Yates, and Xinjie
Wang for fruitful discussions.  This work was supported by NSF Grant No.
DMR 0706493. Computational resources have been provided by NERSC.

\bibliographystyle{apsrev}

\end{document}